\documentstyle[twocolumn,prl,aps]{revtex}
\begin{document}
\draft

\title{Hopf term and the effective Lagrangian for the Skyrmions
in a two--dimensional electron gas at small $g$--factor}

\author{W. Apel$^{1}$ and Yu.~A. Bychkov$^{2,1}$ }

\address{$^{1}$ Physikalisch-Technische Bundesanstalt,
Bundesallee 100, 38116 Braunschweig, Germany.}

\address{$^{2}$ L.D. Landau Institute for Theoretical Physics,
ul.Kosigina, 2, Moscow, Russia.}

\date{\today}
\maketitle

\begin{abstract}
We study interacting electrons in two dimensions moving in the lowest Landau
level under the condition that the Zeeman energy is much smaller than the
Coulomb energy and the filling factor is one.
In this case, Skyrmion quasiparticles play an important role.
Here, we present a simple and transparent derivation of the corresponding
effective Lagrangian.
In its kinetic part, we find a non--zero Hopf term the prefactor of which we
determine rigorously.
In the Hamiltonian part, we calculate, by means of a gradient expansion,
the Skyrmion--Skyrmion interaction completely up to fourth order in spatial
derivatives.
\end{abstract}

\vspace{3ex}


Two--dimensional electron gases as manufactured in GaAs heterostructures show a
rich variety of features as the strength of a magnetic field or the particle
density is varied.  Most prominent among these are the quantum Hall effects
(QHEs) at integer and fractional filling factors $\nu$ ($\nu = N_e/N_{\Phi}$,
where $N_e$ and $N_{\Phi}$ are the number of electrons and the orbital
degeneracy of the Landau level, respectively).
The existence of a well--defined single--particle spectrum of discrete,
spin--split Landau levels allows one to explain the integer QHE.
Until recently it was accepted that when in such a system the characteristic
Coulomb energy is less than the cyclotron energy, the ordinary low--lying
excitations are electron--hole pairs of opposite spins (spin excitons
\cite{bychkov81,kallin84}). These have a non--zero kinetic energy with a strong
$k$--dispersion due to the electron--electron interaction.
In a completely filled Landau level, the energy gap for creating a widely
separated quasielectron--quasihole pair, a large spin exciton (i.e.\ one with
$k \rightarrow \infty$), is apart from the Zeeman splitting governed by the
exchange energy associated with a hole, and is equal to
$|g| \mu_B B + \sqrt{\frac{\pi}{2}} \frac{e^2}{\kappa l_H}$.
Here, $g$ is the effective single particle Land\'e factor and
$l_H = (c\hbar/eB)^{1/2}$ is the magnetic length.
This is in qualitative agreement with experiments on the temperature--dependent
longitudinal resistance in the thermally activated regime
\cite{nicholas88,usher90}.

Recent theoretical investigations \cite{sondhi93,fertig94} near filling factor
$\nu = 1$, however, revealed that the interplay between Zeeman and Coulomb
interaction results in a more complex type of excitations with unusual spin
order which can be described as Skyrmions.
It was shown \cite{sondhi93} that the energy gap required to create a widely
separated Skyrmion--anti-Skyrmion pair, is only half of the gap required to
create a large spin exciton.
Skyrmions appeared originally in Condensed Matter physics as solutions of the
$O(3)$ non--linear sigma model in two dimensions for non--zero values of the
topological charge \cite{belavin75}.
Provided now that the Zeeman energy is less than the Coulomb energy, the
electron gas is equivalent to an isotropic itinerant ferromagnet.
The latter can then be described by a three--component order parameter in a 2D
coordinate space, i.e.\ just by the $O(3)$ non--linear sigma model
\cite{belavin75,rajaraman89,fradkin90} in which one finds topologically
non--trivial spin--textures, the Skyrmions.
Evidence for a realization of these unusual excitations in a 2D electron gas
under a strong magnetic field has recently been uncovered in a number of
experimental papers \cite{barret95,schmeller95,aifer96,maude96} which use
quite different techniques.

Recently, there has also been a lot of theoretical work dealing with the case
of filling factor $\nu$ near and at $1$, i.e.\ with two spin--split Landau
sub--levels; this case is of interest here.
In the problem of a double--layer system, intimately related to our case,
several derivations of a Lagrangian for the Skyrmions have been discussed in
\cite{moon95}.
For finite $|\nu -1|$, a periodic arrangement, a crystal of Skyrmions was
studied \cite{brey95}.
Starting from a Lagrangian description, Ref.\ \cite{read96} discussed finite
temperature properties within a scaling theory.
The many body perturbation theory approach was employed in \cite{kasner96} to
study the thermodynamics.
A different line was followed by one of the present authors \cite{bychkov96}.
It builds on the observation that the Hartree--Fock (HF) approximation
is valid for small $g$--factor, when the Skyrmions contain many reversed spins.
Then, the HF state is parameterized by a unit vector $\vec n(\vec r)$ related
to the spin--density.
The energy is invariant against a uniform rotation of $\vec n$.  Expanding the
HF energy in spatial derivatives leads to a $O(3)$ non--linear sigma model with
a topological term (describing the map $S_2 \rightarrow S_2$).
Here, the latter has a {\it real prefactor}, in remarkable contrast to other
cases of physical realizations of this model, e.g.\ the Heisenberg chain where
an {\it imaginary prefactor} describes the difference between integer and
half--integer spin cf.\ \cite{affleck89,fradkin90}.

There are two very important problems connected with the physics of Skyrmions.
The first regards the interaction between the Skyrmions. In order to describe
this properly, the gradient expansion of the energy functional \cite{bychkov96}
valid in the case of a very small Zeeman energy should be continued at least up
to the fourth order in derivatives.
The second problem lies in the dynamics of the Skyrmions and their spin and
statistics.  These are determined in the Lagrangian by the terms containing time
derivatives.
Now the main theoretical question is how to {\it derive} the corresponding
macroscopic Lagrangian from the microscopic Hamiltonian of interacting electrons
in two spin--split Landau sub--levels.  The particular challenge there regards
the quantum nature of the Skyrmion excitations.
It is determined by the ``Hopf term'' \cite{wilczek83,fradkin90} which is
topological in nature and, obviously, cannot be found by a simple generalization
of the classical model but rather needs to be derived from the microscopic
Hamiltonian of the electrons.
This constitutes the objective of the present work.
So far, it was unclear whether the prefactor of a Hopf term in the action is
non-zero and how it could be found from microscopic calculations.
In this Letter, we set out to give a complete derivation of all parts of
the effective action up to and including the fourth order in derivatives.
This is not a trivial task, since we are expecting terms which are total
derivatives as the Hopf term \cite{wu84} and thus can be easily overlooked.
Our treatment reveals the full $SU(2)$ symmetry of the problem which is
displayed by the kinetic part of the action. The Hamiltonian part, on the
other hand, shows $SU(2)/U(1)$ symmetry.
With our result, the statistics and the interaction of the Skyrmions
are then rigorously determined.
As far as we are aware, this is the first case in Condensed Matter
theory in which one can calculate a non--zero Hopf term from the microscopic
model and thus determine the statistics of the quasiparticles.

We study interacting electrons in the lowest Landau level.
Using the Landau gauge, the destruction operator of a spin up (down) electron
with linear momentum $p$ is denoted by $\hat{a}_p$ ($\hat{b}_p$).
Then, the Hamiltonian $\hat{H}$ reads

\begin{eqnarray}
\hat{H}=\frac{1}{2}\sum_{\vec q,p_1,p_2} &\tilde{V}&(q)
 e^{iq_{x}(p_{2}^{^{\prime}}-p_1)} \,
 [\hat{a}_{p_1}^{\dagger}\hat{a}_{p_2}^{\dagger} \hat{a}_{p_{2}^{^{\prime}}}
 \hat{a}_{p_{1}^{^{\prime}}} + (\hat{a}\rightarrow\hat{b}) \nonumber \\
&+&2\hat{a}_{p_1}^{\dagger}\hat{b}_{p_2}^{\dagger} \hat{b}_{p_{2}^{^{\prime}}}
 \hat{a}_{p_{1}^{^{\prime}}}]  \;.                   \label{hamil}
\end{eqnarray}

Here, as in \cite{bychkov96}, $\tilde{V}(q)=e^{-q^2/2}V(q)$, where
$V(q)$ is the electron--electron interaction, and
$p_{1}^{^{\prime}}=p_1 -q_y$, $p_{2}^{^{\prime}}=p_2 +q_y$.
All lengths are measured here in units of magnetic length $l_{H}$.
We write the HF state as
$|\Psi_{HF}>=\Pi_{p}\hat{A}_{p}^{\dagger}|0>$ (denoted by $|\psi>$ in
\cite{bychkov96}) where the creation operators $\hat{A}_{p}^{\dagger}$
of the single particle HF states are given by

\begin{equation}
\hat{A}_{p}=\sum_{p_1}(U_{p,p_1}\hat{a}_{p_1}+V_{p,p_1}\hat{b}_{p_1}) \;,
\end{equation}

with

\begin{equation}
\hat U = e^{i\hat \psi /2}\cos {\frac{\hat \theta }2}\,e^{i\hat \phi /2} \;,\;\;
\hat V = e^{i\hat \psi /2}\sin {\frac{\hat \theta }2}\,e^{-i\hat \phi /2} \;.
\label{uvtr}
\end{equation}

The matrices $\hat U$ and $\hat V$ represent a transformation
of $\hat{a}_{p}$ and $\hat{b}_{p}$ which respect the anticommutation rules.
They are parameterized by three Euler angles, $\psi(\vec r)$,
$\theta(\vec r)$, and $\phi(\vec r)$ as follows.
The elements $\psi_{p,p^{^{\prime}}}$ of the matrix $\hat \psi$ are to be
calculated with the Landau states from $\psi(\vec r)$ as
$\psi_{p,p^{^{\prime }}}=[\psi (\vec r)]_{p,p^{^{\prime }}}$, and
correspondingly for $\hat \theta$ and $\hat \phi$.
In all the calculations below, we shall use the techniques developed in
\cite{bychkov96}.
For each matrix $\hat M$, define a corresponding function

\begin{equation}
\tilde{M}(\vec q)=\sum_{p}e^{iq_{x}(p+q_{y}/2)} M_{p+q_{y},p} \;.
\label{mt}
\end{equation}

Then, e.g.\

\begin{equation}
\tilde{\theta}(\vec q)  =  e^{- \frac{1}{4} q^2 }
    \int \frac{d^2r}{2 \pi} \, \theta (\vec r) \, e^{-i\vec q\cdot\vec r}
 =: \int \frac{d^2r}{2 \pi} \, \overline{\theta}(\vec r) \,
    e^{-i\vec q\cdot\vec r}  \;.
\end{equation}

$\overline{\theta}(\vec r)$ differs from $\theta(\vec r)$ in that it is averaged
over the area of one flux quantum. Since we are interested in a gradient
expansion in which functions vary slowly on the scale of the magnetic length,
the renormalized $\overline{\theta}$ is the natural variable in which we expect
to express our results.
In the following, we repeatedly need functions of matrices as in (\ref{uvtr}).
With the definition (\ref{mt}) and using the composition rule Eq. (20) in
\cite{bychkov96} we get up to fourth order in gradients e.g.\

\begin{equation}
\int\frac{d^{2}q}{2\pi} e^{i\vec q\cdot\vec r}
\widetilde{ \left[ \cos \frac{\hat \theta }{2} \right] }(\vec q)
= \left( 1 + {\cal O}(\nabla ^4) \right) \;
 \cos \frac{{\overline \theta} (\vec r)}{2}
\end{equation}

and similarly for the other functions in (\ref{uvtr}).

We now proceed with our derivation of the effective Lagrangian.
The HF approximation of $\hat H$ reads \cite{bychkov96}

\begin{eqnarray}
<\hat H>=\sum_{\vec q}\{&\frac 12&[\tilde V(q)-\pi E(q)] \;
          \tilde{N}(\vec q) \; \tilde{N}(-\vec q) \nonumber \\
&-&2\pi E(q) \; \sum_{j=x,y,z} \tilde{S}^j(\vec q) \; \tilde{S}^j(-\vec q) \}
\;. \label{meanham1}
\end{eqnarray}

Here,
$E(q)\equiv\int\frac{d^2 p} {(2\pi)^2}\tilde{V}(p)e^{i\vec p\cdot\vec q}$, and
the matrix corresponding to the HF expectation value of the charge--density
$\tilde{N}(\vec q)$ is given by

\begin{equation}
\hat N = \hat U^{\dagger} \; \hat U + \hat V^{\dagger} \; \hat V \;,
\end{equation}

while the matrices corresponding to the HF expectation values of the vector
components of the spin--density $\tilde{S}^j(\vec q)$ are

\begin{equation}
 {\hat S}^{x} + i {\hat S}^{y} = \hat V^{\dagger} \; \hat U   \;\; , \;\;
 {\hat S}^{z} = \frac{1}{2} \left[ \hat U^{\dagger} \; \hat U -
                  \hat V^{\dagger} \; \hat V \right] \;.
\end{equation}

$\hat N$ and $\hat{S}^j$ and thus (\ref{meanham1}) do not involve the angle
$\overline \psi$ which, therefore, was not taken into account in
\cite{bychkov96}.
Here, it will be necessary to keep the transformation as general as in
(\ref{uvtr}) as will become clear in the sequel.

After having defined the Hamiltonian and the HF state, proceeding now
to evaluating the effective action, we consider time dependent HF states
$|\Psi_{HF}>$ parameterized by $\overline{\psi}(\vec r, t)$,
$\overline{\theta} (\vec r,t)$, and $\overline{\phi} (\vec r, t)$.
Then, our task is to expand the effective Lagrangian (real time)

\begin{equation}
{\cal L} = <\Psi_{HF}| i {\partial_t} |\Psi_{HF}>
          - <\Psi_{HF}| \hat H |\Psi_{HF}>
\label{lagr}
\end{equation}

with respect to gradients in the fields $\overline \psi$, $\overline \theta$,
and $\overline \phi$ up to and including fourth order.\\[1ex]

{\it Kinetic part}\hspace{1em}
In calculating

\begin{eqnarray}
{\cal L}_k &=& <\Psi_{HF}| i\partial_t |\Psi_{HF}>
= \sum_{p} <0|\hat{A}_{p} \, i\partial_t \hat{A}_{p}^{\dagger}|0> \nonumber \\
&=& \sum_{p,p_1} \left\{ U_{p,p_1} \, i\partial_t U_{p_1,p}^{\dagger} +
                    V_{p,p_1} \, i\partial_t V_{p_1,p}^{\dagger} \right\} \;,
\end{eqnarray}

there is a delicate point. After taking the derivatives, one should not use such
properties of the trace as cyclic permutations, since the sums in question do
not converge absolutely.  Disregarding this, one would miss total derivatives
and hence an important part of the result.  Consequently, we first get the
function ${\tilde U}(\vec q,t)$ corresponding to $U_{p,p_1}$ etc.\ and then
calculate

\begin{equation}
{\cal L}_k = \int\frac{d^{2}q}{2\pi}
\left\{ \tilde{U}(-\vec q,t) \, i\partial_t \tilde{U^{\dagger}}(\vec q,t)  +
        ( U \rightarrow V ) \right\} \;.
\end{equation}

Using again the composition rule Eq.\ (20) in \cite{bychkov96} we find
up to fourth order in gradients ${\cal L}_k = {\cal L}_k^0 + {\cal L}_k^{top}$
(a spurious imaginary term in ${\cal L}_k$ turns out to be the time derivative
of the topological charge and hence vanishes \cite{apel96}),

\begin{equation}
{\cal L}_k^0 = \frac{1}{4\pi} \int d^{2}r
 \left[ \partial_t \overline{\psi}(\vec r,t) + \cos \overline{\theta}(\vec r,t)
 \, \partial_t \overline{\phi}(\vec r,t) \right]
\end{equation}

and

\begin{equation}
{\cal L}_k^{top} = \frac{-1}{16\pi} \int d^{2}r  \left\{
  \frac{\partial (\overline{\psi},\cos \overline{\theta},\overline{\phi})}
       {\partial (t,x,y)}
- \partial_t \left[ \cos \overline{\theta} \,
          \frac{\partial (\overline{\phi},\overline{\psi})}{\partial (x,y)}
             \right] \right\}  \;.           \label{l2}
\end{equation}

Only the part ${\cal L}_k^0$ of ${\cal L}_k$ contributes to the equations of
motion, since ${\cal L}_k^{top}$ is a total derivative in the action.
As expected, we find from ${\cal L}_k^0$ the kinetics of a system of spins
$\frac{1}{2}$.
${\cal L}_k^0$ can be expressed in several ways; here it is given in a
form explicitely depending on our representation of the unit vector $\vec n$

\begin{equation}
\vec n \equiv
  (\sin{\overline{\theta}} \cos{\overline{\phi}},
   \sin{\overline{\theta}} \sin{\overline{\phi}},
   \cos{\overline{\theta}} )
\end{equation}

which describes the spin in terms of the renormalized angles
$\overline{\theta}$ and $\overline{\phi}$ cf.\ \cite{klauder79}.
The $\overline \psi$ degree of freedom decouples in this part of the action.
Since it also does not enter the definition of the spin-- and charge--density,
cf.\ \cite{bychkov96}, one would have a model with symmetry $SU(2)/U(1)$ for
$\vec n$, if it were not for the second part ${\cal L}_k^{top}$, which we
derived here for the first time.

${\cal L}_k^{top}$ contains firstly a non-zero ``Hopf term'' and also an
additional total derivative in time.
The Hopf term is the linking coefficient of a smooth map $S_3 \rightarrow S_2$
\cite{dubrovin86}.
Usually, it is expressed as \cite{wu84}

\begin{equation}
 {\cal H}_{Hopf} = \frac{\Theta}{4\pi^2} \int d^2r dt \,
            \epsilon^{\mu \nu \lambda} \, (z^{*} \partial_{\mu} z) \,
            (\partial_{\nu} z^{*} \partial_{\lambda} z) \;.
\end{equation}

Taking for the spinor $z$

\begin{equation}
 z = ( e^{\frac{i}{2}(\overline\psi-\overline\phi)}
          \cos \frac{\overline\theta}{2},
       e^{\frac{i}{2}(\overline\psi+\overline\phi)}
          \sin \frac{\overline\theta}{2} ) \;,
\end{equation}

then the unit vector $\vec n=z^{*} {\vec \sigma} z$ ($\sigma^j$ are the Pauli
matrices), and we identify ${\cal H}_{Hopf}$ with the first term in brackets of
our result (\ref{l2}).
Its prefactor is such that the $\Theta$--angle describing the statistics
of the Skyrmions in Ref.\ \cite{wilczek83} is equal to $\pi$.
Thus, in the language of \cite{wilczek83}, the skyrmions carry spin
$\frac{1}{2}$.
The microscopic derivation of this is the first of the main results of
the present paper.
The additional term in (\ref{l2}) finds a quite natural explanation: it
enters in such a way that this part of the density of the effective Lagrangian
becomes a {\it total spatial gradient} (still including a time derivative)
without any total time derivative.\\[1ex]

{\it Hamiltonian part}\hspace{1em}
Next we proceed to calculate the second part in the effective Lagrangian
(\ref{lagr}).
Using the methods described above, $\tilde{S}^j(\vec q)$ and
$\tilde{ N}(\vec q) = 2\pi \delta (\vec q) + \delta N(\vec q)$ are
expanded in gradients of the fields and also $E(q)$ is expanded in $q$ up to
fourth order.
(We omit the time--argument which is unnecessary in this part.)
In lowest order, the spin--density in real space is given by

\begin{equation}
\tilde{S}^j(\vec r) = \frac{1}{4\pi}  n^j(\vec r) \;.
\end{equation}

From $\vec n$, the charge--density is determined as

\begin{equation}
\delta N(\vec r) = \frac{1}{4\pi} \vec n(\vec r) \cdot
               \partial_x \vec n(\vec r) \times \partial_y \vec n(\vec r)  \;.
\label{numpar}
\end{equation}

The next order in the gradient expansion, i.e.\ second for $\tilde{S}^j$ and
fourth for $\tilde N$ vanishes and one is left with the calculation of the
fourth order for $\tilde{S}^j$.  Collecting all terms together, we find
$<\Psi_{HF}| \hat H |\Psi_{HF}> = {\cal L}_H + {\cal O}(\nabla ^6)$
\cite{apel96}

\begin{eqnarray}
{\cal L}_H &=& \frac{E(0)}{8\pi} \int d^{2}r \left\{ \frac{1}{4}
               \sum_{\alpha=x,y} (\partial_{\alpha} \; \vec n)^{2}
               -  \vec n \cdot
                   \partial_x \vec n \times \partial_y \vec n \right\}
                                                                   \nonumber \\
               &-& \frac{3E(0)}{2^9\pi} \int d^{2}r (\triangle \vec n)^2
               + \frac{1}{2}\sum_{\vec q}\tilde{V}(q) |\delta N(\vec q)|^2 \;.
\label{lh}
\end{eqnarray}

Eq. (\ref{lh}) is the second main result of the present paper.
The terms of second order in derivatives in (\ref{lh}) represent a $O(3)$
non--linear sigma model with topological term with a {\it real prefactor}.
For a discussion of these second order terms in (\ref{lh}), cf.\
\cite{bychkov96}.
As regards the next order, one could have written down a number of terms of
fourth order in derivatives of the components of $\vec n$ which all show the
required symmetries.
Surprisingly however, apart from the Coulomb energy which had to come
out, there is basically no correction to the lower order calculation
in which the Skyrmions are still non--interacting.
One gets only the term proportional to $\triangle ^2$ which comes from
the direct expansion of $E(q)$ (its numerical prefactor is here given for the
Coulomb interaction and the lowest Landau level).
This term is only a small correction in our expansion in the variation of the
fields on the scale of the magnetic length.
A simple estimate shows that it contributes a term of the order of
$-\frac{e^2}{\kappa l_H} (l_H/r_0)^4$ ($r_0$ is the size of the Skyrmion)
to the energy of a single Skyrmion and thus does not help to stabilize the
Skyrmion for small values of the $g$--factor.
In (\ref{lh}), a total derivative in spatial coordinates was omitted
which in all cases we checked was non--singular and thus did not
contribute to (\ref{lh}).\\[3ex]

We thank M.\ Potemski for sharing the results of \cite{maude96} with us prior to
publication.
One of us (Yu.A. B.) thanks the PTB for hospitality and
acknowledges support of the INTAS-94-4055 and the RFFI 95-02-05745 projects.


\begin{thebibliography}{99}
\bibitem{apel96}  A detailed account of this work will be published later.
\end{thebibliography}
\end{document}